\documentclass{article}
\usepackage{epsfig}
\usepackage{booktabs}
\usepackage{ucs}
\usepackage{url}
\usepackage[utf8x]{inputenc}
\usepackage{xcolor}
\usepackage{enumitem}

\title{\textbf{Automatic Bias Detection in Source Code Review}}

\author{
\begin{tabular}{p{0.45\linewidth} p{0.45\linewidth}}
\multicolumn{1}{c}{\textbf{Yoseph Berhanu Alebachew}} & \multicolumn{1}{c}{\textbf{Chris Brown}} \\
\multicolumn{1}{c}{Department of Computer Science} & \multicolumn{1}{c}{Department of Computer Science} \\
\multicolumn{1}{c}{Virginia Tech} & \multicolumn{1}{c}{Virginia Tech} \\
\multicolumn{1}{c}{\texttt{yoseph@vt.edu}} & \multicolumn{1}{c}{\texttt{dcbrown@vt.edu}} \\
\end{tabular}
}

\date{September 9, 2024}

\date{Sept 6,2024}

\begin{document}

\maketitle

\begin{abstract}
Bias is an inherent threat to human decision-making, including in decisions made during software development. Extensive research has demonstrated the presence of biases at various stages of the software development life-cycle. Notably, code reviews are highly susceptible to prejudice-induced biases, and individuals are often unaware of these biases as they occur. Developing methods to automatically detect these biases is crucial for addressing the associated challenges. Recent advancements in visual data analytics have shown promising results in detecting potential biases by analyzing user interaction patterns. In this project, we propose a controlled experiment to extend this approach to detect potentially biased outcomes in code reviews by observing how reviewers interact with the code. We employ the "spotlight model of attention", a cognitive framework where a reviewer’s gaze is tracked to determine their focus areas on the review screen. This focus, identified through gaze tracking, serves as an indicator of the reviewer's areas of interest or concern. We plan to analyze the sequence of gaze focus using advanced sequence modeling techniques, including Markov Models, Recurrent Neural Networks (RNNs), and Conditional Random Fields (CRF). These techniques will help us identify patterns that may suggest biased interactions. We anticipate that the ability to automatically detect potentially biased interactions in code reviews will significantly reduce unnecessary push-backs, enhance operational efficiency, and foster greater diversity and inclusion in software development. This approach not only helps in identifying biases but also in creating a more equitable development environment by mitigating these biases effectively.
\end{abstract}

\section{Introduction}
The Software Development Life Cycle (SDLC) is full of decisions. For example, when working in a team, developers perform an activity called code review which requires the reviewer deliberate on whether to include the code written by other developers into the main code-base. In open source project that utilize distributed version control systems such as Git\footnote{\url{https://git-scm.com/}}, this review process will take a from of pull request. 

Like any human decision-making process, stakeholders in the SDLC are susceptible to cognitive bias. In a mapping study that assessed over 65 scientific papers published up to 2016, Mohanani \textit{et al.} \cite{mohanani2020cognitive} reported that at least 37 cognitive biases have been shown to exist in the software engineering process. Based on a study that observed 16 developers in situ, Chattopadhyay \textit{et al.} \cite{chattopadhyay2022cognitive} reported that 45.72\% of developers’ actions exhibited one or more of 28 biases.

In a large-scale pull request analysis, Murphy-Hill \textit{et al.} \cite{murphy2022pushback} discovered that biased pushbacks exist with a high correlation to the code author's gender, ethnicity, and age. The authors estimated that over 1000 hours of programmer time are wasted per day due to this pushback. Such push back could be a result of in-group favoritism, which is when one give preferential treatment to others they perceive as being part of their own group, whether that group is defined by ethnicity, nationality, religion, or other shared characteristics.

With the pervasiveness of open source software projects and global software development powered by distributed version control systems such as git, programmers work with code contributors/authors of different background. This means, the source code review process (often in the form of pull/merge request) involves accepting/rejecting code written by a contributor whom the reviewer might no know at all. Platforms like GitHub\footnote{\url{https://github.com/}} provide some profile information about the code author. This in principle should not be a problem as code review should only involve examining the source code and decide on it's merit only. However, it has been shown that reviewers examine information besides the source code while performing the review \cite{ford2019beyond}. In the study participants reviewed source code while their eye gaze was tracked. This authors reported that reviewers gazed and fixated on profile indicators such as profile picture of the authors more than what was self reported by the participants.

In a follow-up to the aforementioned study, Huang \textit{et. al}\cite{huang2020biases}  established a causation relationship between reported author gender and biased merge request decision. In this study, participants presented with pull request were more likely reject it if the author was reported to be a female or an automated bug fix program than if the same pull request was presented as being authored by male contributor.  
\subsection{Background}
Decision-making refers to the process of choosing among several alternatives. We typically associate decision-making with instances where we make conscious and deliberate choices. However, based on this definition, we make decisions every minute of every day, often with little conscious effort. Generally, we want to and believe we make rational decisions, that is, we choose the alternative that is most likely to lead to our desired outcome \cite{Pronin2002}. 

Making such rational decisions, however, requires that we weigh all alternatives and the potential outcomes for each, which is resource-intensive. There is also a trade-off between the accuracy and timeliness of our decisions. As a result, we tend to take shortcuts called heuristics. These short cuts enable us to make decisions that, while not guaranteed to be rational all the time, work often enough and are far timelier than exploring all alternatives in detail. These shortcuts take various forms but generally result from an effort to achieve decision-making goals with minimal resource expenditure.

When these shortcuts fail to result in a rational outcome, it leads to what we call cognitive bias. Tversky and Kahneman, who first introduced the concept of cognitive bias, defined it as "the systematic errors in judgment that occur predictably in particular circumstances" \cite{tversky1974judgment}. One key aspect of this definition is that biases are predictable. This means that under the right circumstances, the likelihood of making a cognitive bias is more than random chance. Since the introduction of the concept in 1974, over 200 cognitive biases have been identified by researchers in behavioral economics, psychology, management, and sociology.

\subsection{Problem Statement}
The core problem we aim to address in this proposed project is the need for an automated system to detect bias in source code reviews. Bias in code reviews can significantly impact the quality and fairness of the review process, leading to potential issues in code quality, team dynamics, and overall project outcomes. Currently, detecting such biases is challenging due to the subtle and often unconscious nature of biased behavior. Our approach involves analyzing data on how reviewers interact with the code review screen. By examining patterns in actions such as scrolling, highlighting, commenting, and time spent on different sections of the code, we aim to identify indicators of bias. This attempt to identify patterns of interaction that can indicate possibility of bias will provide valuable insights into the review process, enabling teams to address and mitigate bias, thereby improving the fairness and effectiveness of code reviews.

\section{Literature Review}
\subsection{Physical Manifestation of Bias}
Boonprakong \textit{et al.} \cite{boonprakong2023bias} explored the indicators of bias in physiological and interaction data. They demonstrated the existence of behavioral and physiological signals of cognitive bias. The authors conducted two studies that collected and analyzed eye-tracking, skin conductance level (SCL), and fNIRS data from an experiment in which participants were presented with information that either aligned with or opposed their opinions on a topic. Since the opinions were presented in text form, the eye-tracking data collected focused on dwelling time instead of scan path.

Using the data from the experiment, the authors built four bias classifiers: a Linear Discriminant Analysis (LDA), Support Vector Machine (SVM) with RBF kernel, Random Forest, and XGBoost, all with 5-fold cross-validation. The highest accuracy achieved was 55.27\% with XGBoost, which barely outperformed ZeroR (50.04\%). To ensure that the models' performance was not due to mere chance, they conducted permutation tests and found that all achieved a p-value lower than 0.05. Participants showed higher neural activity and spent more time processing statements that disagreed with their own opinions.

\subsection{Cognitive Bias in Software Engineering}
The Software Engineering (SE) process is intricate and decision-intensive. Like any decision-making endeavor, SE is prone to cognitive biases. Mohanani \textit{et al.} \cite{mohanani2020cognitive} conducted a comprehensive mapping study to explore cognitive biases within the field of software engineering. This study aimed to identify, catalog, summarize, and synthesize existing research on the subject. Specifically, it highlighted research on cognitive biases frequently encountered in SE, focusing on their origins, manifestations, and impacts. The authors examined 65 papers published until 2016, finding that 37 cognitive biases have been studied at various stages of the SE process.

Chattopadhyay \textit{et al.}\cite{chattopadhyay2022cognitive} conducted a two-part field study to assess the extent of cognitive biases in developers' daily activities. The field study took place in a startup, where the authors observed ten developers in situ for an hour while they performed their daily tasks. They found that 45.72\% of developers’ actions (953 out of 2084) involved one or more of the 28 observed biases, which were grouped into ten categories.

As part of the study, follow-up interviews were conducted with 16 developers to validate the findings, assess the perceived frequency of biases, and identify available tools and practices to address the identified biases as reported by developers. The authors reported that developers spent 34.51\% of their time reversing biased actions. Of all the reversal actions, 70.07\% involved biases. They defined reversal actions as those that developers need to undo, redo, or discard and measured the total number of actions and time spent on these reversal actions resulting from cognitive biases.

Additionally, the authors assessed the standard tools developers reported for bias mitigation and presented a set of helpful practices to mitigate or address the biases, grouped into six categories.

\subsection{Bias in Source Code Review}
One area within software engineering where biases have been observed is the process of source code review. With the widespread adoption of distributed version control systems, code reviews often occur through merge requests (or change requests). Murphy-Hill \textit{et al.} \cite{murphy2022pushback} conducted a large-scale analysis of pull requests at Google. The study's primary objective was to examine the influence of demographic attributes, such as gender, ethnicity, and age, on the rate of pushback. Here, "pushback" is defined as "the perception of unnecessary interpersonal conflict in code review, where a reviewer blocks a change request" \cite{egelman2020predicting}. This concept draws on role congruity theory, which posits that a group member receives negative evaluations when stereotypes about the group are misaligned with the qualities perceived as necessary for success in a specific role. The authors found a significant correlation between the incidence of pushback and the gender, ethnicity, and age of the code's author.

An eye tracking experiment by Ford \textit{et. al} \cite{ford2019beyond} has showed that reviewers do look at information beyond the source code itself while decision to accept or reject a pull request. Reviewers fixated on what Ford \textit{et. al.} referred to as \textit{social indicators} such as the contributor’s aviator, display name, and contribution history. They found that the amount of actual fixation was more than what was self-reported by reviewers after the experiment.

In another eye tracking study, Huang \textit{et. al} \cite{huang2020biases} re-established the existence of correlation reported gender of authors and rate of push back. What's more they  demonstrated causation relationship exists. They found both male and female reviewers were harsher on female-authored pull requests compared to male contributors of the same contribution. This implies the reviewer's bias based on gender which contradicts the self-reported behavior in which the reviewers claimed they did not put the gender of the author into consideration.

An interesting finding in this study is that participants admitted their bias towards computer generated code. But even in these instances participants tried to rationalize their bias after the fact with excuses such as that the code generated by computer programs are "too complex and less understandable".This is despite the fact that the actual code was written by human. This shows that even when we acknowledge our biases, we try to justify them rather than gauge their rationality. 

\subsection{Bias Detection}
Detecting cognitive bias refers to identifying the expected likelihood of bias in a decision rather than certainty. Primary procedure for detection has been based on selective exposure experiment. In such methods, participants are surveyed with pre and post experiment questionnaires which will be used as a basis to see if they are biased in the task. Such an approach can be referred to as a static detection since it relies on post activity static analysis as opposed to dynamic techniques that try to detect bias during the decision making. Static detection limits the debiasing techniques that can be employed.

Dynamic Detection is a detection effort whereby the (possible) occurrence of cognitive bias is detected during the deliberation process, before the actual decision is made. While such detection is more likely to help in mitigating biases in the current decision, it is relatively difficult to implement. 

Wall \textit{et. al.} \cite{wall2017warning} proposed a theoretical framework for quantifying interactions and subsequently identifying biased interaction in visual data analytics tools. In a later work \cite{wall2019formative}, the authors presented a formative study that implemented this theoretical recommendation. They used their proposed metrics in detecting anchoring bias from interaction data while participants were tasked with an activity purposefully prepared to induce bias. The authors reported an encouraging result towards the ability to detect cognitive bias, in real time, based on interaction behavior.

Nussbaumer et al. \cite{nussbaumer2016framework} presented an attempt to establish a framework for automatic confirmation bias detection and feedback in criminal intelligence visual analytics. The authors identified software development and research challenges that need to be addressed to realize the proposed framework. The research tasks include developing a detection method, providing meaningful feedback, and evaluating the effectiveness of the proposed methods.

They proposed the use of visual analytics tools, such as MUVA \cite{Kalamaras2015-br} and VALCRI\footnote{https://cordis.europa.eu/project/id/608142}, to record interaction logs for analysis while users performed typical tasks. This work defines two types of bias detection based on interaction data. The first is a statistical approach that compares biased interactions with unbiased ones. The second is a semantic approach that maps biased and unbiased interactions to cognitive processes and compares these cognitive processes.


\subsection{Eye Tracking in Software Engineering}
Eye tracking has been used by a number of researchers to understand developer behavior. Bansal et al. \cite{bansal2023modeling} present an early attempt at using Large Language Models (LLMs) for programmer attention modeling. They employed gaze trackers and LLMs to build a machine learning model capable of predicting how programmers scan source code. To build this model, the authors had 27 programmers perform a Java source code comprehension task, involving 25 randomly selected methods from a total of 68, for 1.5 hours with breaks every 20 minutes. The participants wore eye trackers to record their fixations during the task. Afterward, the participants were asked to write short summaries of the source code. The collected data was used to build an LLM model that takes source code as input and produces the expected scan-path. This model aims to mimic the way programmers would scan and comprehend source code, providing insights into programmer attention patterns.

Aljehane et al. \cite{aljehane2023studying} used eye movement data to identify the expertise level of developers based on how they examine source code. They conducted a quantitative analysis comparing the eye movements of 207 participants, including both novices and experts, while they solved comprehension tasks. The results revealed a notable increase in pupil size among the novice group compared to the experts, suggesting that novices exert greater cognitive effort. Additionally, novices exhibited significantly more fixations and longer gaze durations than experts when comprehending code. Furthermore, a correlation study indicated that programming experience remains a strong predictor of expertise in this eye-tracking dataset, alongside other expertise variables.

\section{Proposed Methodology}
We propose a research project aimed at collecting data on code reviewer interactions through an experimental setup. We will analyze the collected data using various machine learning algorithms to identify patterns that predict the occurrence of cognitive biases, with a particular focus on prejudice. Our findings will be reported at a peer-reviewed academic conference. This section details the major tasks envisioned for our project.
\subsection{Research Design and Data Collection}
Our plan involves conducting an experiment to measure participants' performance in source code review tasks. We will first identify open-source pull requests suitable for the experiment based on criteria set by related research works. Then, we will recruit participants and ensure they have the required background to review the code under consideration through a survey. We expect to include both industry professionals and experienced students as developers in our study. Participants will be seated in front of computers and presented with a sequence of pull requests, deciding whether to accept or reject each. Code snippets from pull requests will be randomly assigned labels indicating authorship by different groups of contributors.

We plan to use the Tobii Pro Fusion Eye Tracker \footnote{https://www.tobii.com/products/eye-trackers/screen-based/tobii-pro-fusion}  to track participants' gaze. Additionally, we intend to record additional signals to explore the possibility of multimodal analysis including keyboard log, cursor movement and comments/feedback on code and decision outcome. To avoid introducing social desirability bias \cite{Grimm2010-cu} by participants, we plan to carefully design our pre-experiment briefing. We will ensure that no mention of bias is made during the initial briefing to prevent influencing participants' behavior. This approach will help us obtain more genuine responses and interactions during the source code review tasks.

Once the experiment is conducted, we will conduct a thorough debriefing session with each participant in the form of an interview. During this debriefing, we will explain the true purpose of the study and the reason for withholding information about bias in the initial briefing. We will discuss the concept of social desirability bias and how it could have impacted their behavior if they had been aware of it from the start.

After explaining the rationale behind the deception, we will seek the participants' consent post hoc. This means obtaining their permission to use the data collected during the experiment, now that they are fully informed about the study's aims and methods. This process ensures ethical transparency and respects participants' autonomy, as they can choose whether to allow their data to be used in the study after being fully informed. By implementing this strategy, we aim to gather more accurate and reliable data while maintaining ethical standards in our research. We will obtain approval from our local institutional review board (IRB) on our research methods before recruiting participants and commencing this study. 

\subsection{Data Analysis}
After collecting the necessary data, we aim to explore various machine learning models to find the most effective approaches for bias detection and prediction. We will start with simpler probabilistic models, such as those proposed by Wall \textit{et al.} \cite{wall2019formative}, and gradually move towards designing more complex multi-modal and transformer-based deep learning models.

Initially, we will implement and evaluate probabilistic models to establish a baseline performance for bias detection. These models are known for their simplicity and interpretability, which will provide valuable insights into the basic patterns and correlations present in our data.

Next, we will design and evaluate multi-modal machine learning models that can leverage the diverse types of data we are collecting, including eye-tracking data, video recordings, and survey responses. Given the multi-modal nature of our data, we anticipate that these models will outperform simpler, single-modality machine learning algorithms, such as Markov Chains. The integration of different data modalities is expected to capture more nuanced patterns of bias and provide a richer representation of the participants' cognitive processes.

We will also explore the use of advanced deep learning models, such as transformer-based architectures, which have shown significant promise in handling complex, high-dimensional data. These models, with their ability to process sequential and multi-modal data effectively, will be crucial in capturing the temporal and contextual dependencies inherent in our data.

In addition, we will evaluate the performance of various machine learning models, such as Recurrent Neural Networks (RNNs) and Conditional Random Fields (CRF), on the bias prediction task. RNNs are particularly well-suited for sequential data and can capture temporal dependencies, while CRFs are effective for structured prediction tasks, making them valuable for detecting patterns of bias in our data.

Beyond constructing and evaluating machine learning models, we intend to delve into psychological theories that explain human decision-making processes. This interdisciplinary approach will help us understand the underlying mechanisms of cognitive biases and provide a theoretical foundation for interpreting our findings. By integrating insights from psychology, we aim to develop more robust and explainable models that not only predict bias but also offer explanations for why certain patterns of bias occur.

Ultimately, our goal is to develop a comprehensive framework for bias detection and prediction that combines the strengths of machine learning with psychological theories. This framework will enable us to identify and mitigate biases more effectively, contributing to the broader field of cognitive bias research and its applications in various domains.

\section{Expected Outcomes and Future Work}
Upon successful completion of our study, we expect to have:
\begin{itemize}[noitemsep]
    \item conducted a user study to analyze developers' gaze while reviewing code;
    \item collected insights from developers on in-group favoritism bias in code review processes; and
    \item developed a machine learning model capable of predicting the likelihood of bias in code review based on how the reviewer interacts with the code.
\end{itemize} 

The resulting model will leverage the comprehensive multi-modal data we collect, including eye-tracking data, video recordings of facial expressions, and detailed interaction logs, to identify subtle patterns indicative of bias. By analyzing these interaction patterns, our model will be able to provide real-time feedback to reviewers and highlighting potential biases. This will not only enhance the fairness and accuracy of code reviews but also contribute to a more objective and inclusive review process.


The successful implementation of our study will result in a machine learning model that can predict the possibility of bias in code reviews, backed by rigorous data analysis. This model has the potential to help in building a tool that significantly improves the quality and impartiality of software development practices. Such a tool and a more in-depth theoretical interpretation of our finding will be possible extensions of the envisioned study. Once possible occurrence of bias is predicted what to do with the information and how or when to present this information will another question to investigate.
\section{Conclusion}
This proposal highlights the issue of cognitive bias in source code reviews, affecting software development fairness and effectiveness. Research shows biases based on gender, ethnicity, and age can negatively impact review outcomes. We propose using multi-modal machine learning to address this problem. By analyzing eye-tracking data, video recordings, and interaction logs, we aim to detect bias patterns.

Our experimental design involves recruiting participants to review open-source pull requests, using the Tobii Pro Fusion Eye Tracker, and recording additional signals for multimodal analysis. Pre-experiment briefings and thorough debriefings will ensure ethical transparency and participant consent.

We will explore various machine learning models to develop robust models that predict bias and can be studied to offer insights into cognitive processes. Upon completion, we expect to develop a model for predicting and addressing bias in code reviews. This can be used in creating training programs and tools to help developers recognize and counteract their biases. This research aims to improve software development practices, fostering a more equitable and efficient environment.

\end{document}